# The Formation of Thermals on the Ground [*)]

Oliver Predelli
Braunschweig, Germany
Oliver@gliderpilot.eu

## Abstract

The measurement data from the boundary layer measurement masts in Billwerder and Hyytiälä make thermals visible. For this, the temperature and humidity data must first be transformed to be independent of altitude. The subsequent visualisation of this data shows an impressive cross-section of the lower atmosphere with individual updrafts, but also with air bubbles detaching from the ground or falling from higher altitudes. These images give glider pilots in particular a better understanding of how thermals are formed near the ground.

**Keywords:** Convective Boundary Layer · Thermals · Temperature · Humidity

## Introduction

A transmission mast of the NDR (a German radio and television station) is located in Hamburg-Billwerder [CEN1]. The 304-metre-high mast has been equipped with measuring devices at various heights by the Meteorological Institute of the University of Hamburg. When I asked the institute if I could get data, they made the entire measurement campaign available to me from May 2019 [CEN2]. The "boundary layer measurement mast" offers ideal conditions for studying thermals near the ground.

Billwerder data makes thermals visible! The mast is equipped with various weather sensors mounted on small platforms on the outside at different heights. Relevant for the evaluations are the air pressure at a height of 2 metres as well as the air temperature and relative humidity at heights of 2, 10, 50, 110, 175, 250 and 280 metres. With these data, a vertical temperature and humidity cross-section can be drawn through the lower air layer every minute. For this article, a few illustrative examples were selected from the large amount of data. The images show phenomena that can be derived directly from the measured values. This article does not consider what happens to thermals at higher altitudes or in other regions.

## Data analysis and visualisation

One can do little with the pure measured values. In **Fig. 1**, the red lines have the same temperature, the white lines the same air density, and the yellow-green-blue background describes the relative humidity. At about 12:10 UTC it started to rain. For about 10 minutes, the high humidity seems to fall "from above", with temperatures dropping and air density increasing slightly. Later, around 12:40 UTC, the air becomes more and more turbulent. Temperature and density fluctuate strongly, and something is happening with the humidity in the background. Above 150 m, for a few minutes the air temperature is warmer and the air density is slightly lighter than usual. This could be a detachment. But it's hard to tell from the picture. You have to be an experienced meteorologist to interpret this raw data.

The problem with the raw data is the influence of altitude (more precisely of air pressure) on air temperature, density and humidity. It is common knowledge that air temperature decreases with increasing altitude, so it is difficult to compare a measurement at 280 m altitude with a measurement at 50 m altitude. If one could remove the influence of altitude from the data, the pictures would be much clearer. Meteorologists use a trick in such cases by introducing the "potential temperature". This is the fictitious temperature that an air parcel would have if it were adiabatically brought to the altitude of 1000 hPa. In the same way, one can also calculate a "potential air density", i.e. the density that the air parcel would have if it were brought to 1000 hPa. This makes temperature and density a standard of comparison independent of altitude. Temperature and density are thus "projected" onto the pressure surface of 1000 hPa.

As glider pilots we know the "dry-adiabatic temperature gradient", where the air temperature drops by about -1 °C per 100 metres. The thermal always follows exactly this temperature gradient during the ascent. The potential temperature is found where the temperature line intersects with the altitude of 1000 hPa. All air temperatures along the line of the same dry-adiabatic temperature gradient have therefore the same potential temperature. This means nothing else than that thermals always have the same potential temperature, no matter at what altitude.

Now we still need the relative humidity. At this point, the height-independent "mixing ratio" comes into play. The mixing ratio describes the grams of water contained in one kilogram of dry air. The conversion of air temperature, density and humidity into potential temperature, potential density and mixing ratio is done with well-known meteorological equations, which I will not go into here.

The surprising result of this transformation can be seen in **Fig. 2**. Suddenly, a clearly recognisable updraft peels out of the wavy lines of Fig. 1! Even individual "bubbles" within the updraft can be seen.

A few minutes after the passage of a light shower event, heat movement occurs in the atmosphere. The temperature on the ground rises by more than one degree. The restlessness in the air is remarkable. It bubbles and billows incessantly. Sometimes it looks like an updraft is forming, but then it doesn't make it and falls back again. It is far from calmly warming up on the ground, and then at some point - plop - a thermal bubble is released.

To understand the difference between Fig. 1 and Fig. 2, consider two temperature values. In Fig. 1, 10 °C is measured at 50 m altitude at 12:30 UTC and 7.5 °C at 280 m altitude at 12:40 UTC. One cannot even assume that the two temperatures are directly related. However, it can be seen from Fig. 2 that the corresponding potential temperatures at 9 °C are identical. Moreover, these potential 9 °C are lifted by thermals at 50 m altitude at 12:30 UTC and reach an altitude of 280 m 10 minutes later. The air parcel at 280 m at 12:40 UTC is obviously the same as at 50 m at 12:30 UTC. Where else would the warm air at this altitude come from if not "from below"? This cannot be seen in Fig. 1, only in Fig. 2.

What moves upwards out of the picture at about 12:40 UTC is nothing other than an updraft. When lines of the same potential temperature leave the over-adiabatic layer, which is about 50 m thick here, and rise to greater heights, and when the potential densities drop in parallel, this is an unmistakable sign of a detachment. The updraft builds up and will keep the same potential temperature over its entire height. The gradient of this temperature rise also indicates the strength of the thermal. At 12:38 UTC, the temperature line of 9° C rises from 200 m to 280 m within two minutes, corresponding to an updraft of 0.7 m/s.

The rise of the 8.7 °C line at about 12:20 UTC is probably not yet a thermal. During the rain shower, the air was cooled considerably and is now returning to its previous values with a potential temperature of 8.7 °C at about 280 m altitude. This temperature increase could simply be the result of wind blowing the "normally warm" air from the side.

## Details on thermal formation over grassland

What exactly happens during the detachment? To find out, we "zoom in" on the image (Fig. 3). Here we come across very interesting details that show the core of the thermal. Within the thermal, bubbles rise that are themselves warmer than the actual thermal tube. The bubbles separate from the warm air near the ground. This repeats every few minutes. In the process, the air near the ground warms up because the ground, which is shone on by the sun, gives off heat to it. However, this warm air does not rise immediately. It remains on the ground for several minutes. At the same time, solar radiation leads to the evaporation of soil moisture. The local mixing ratio, a measure of air humidity, rises. Only when the humidity has risen sufficiently the bubble takes off. This is a typical behaviour over grassy areas that can also be observed on other days.

It is obvious that a rise in temperature alone is not sufficient to cause detachment. In the May data, it is extremely rare that either only the temperature or only the mixing ratio increases. Usually, temperature and humidity increases occur together. This means that a thermal bubble always has a temperature and humidity advantage over the ambient air when it leaves the overadiabatic layer. Both provide a lower air density within the thermal and drive its buoyancy. Normally, a detachment is about 0.3 to 0.8 °C warmer than the ambient air and its mixing ratio is between 0.1 and 0.5 g/kg higher.

By the way: The detachment in Fig. 3 is triggered from the outside. In fact, it's not like it detaches by itself. Not every bubble you see in the picture rises to the top. There are also bubbles with lower temperature and humidity that fall from the top due to their higher density (at 12:30 UTC). These bubbles displace the warm, moist air at the ground and lift it. This "kick" is enough to separate the air from the ground and start the thermal.

## Thermal formation over forests

Thermals over forests are different. In Finland, the University of Helsinki maintains a measurement mast in Hyytiälä, about 60 km northeast of Tampere [HYY1]. It is part of a research station in the middle of a dense pine forest near a lake. With 125 m, this mast is not as high as the mast in Billwerder. But its nine measurement levels allow a much higher resolution of temperature and humidity. The measurement data of "SMEAR II Hyytiälä forest" is freely accessible on the internet [HYY2].

**Fig. 4** shows the values on a warm day in August. Two effects are clearly different from Billwerder. Firstly, the air between the trees is relatively humid. This is because, in contrast to the grasses in Billwerder, the trees transpire much more and release significantly more water into the air. A second zone of humid air forms above the trees, reaching up to about 60 metres. It is rare to find such a uniformly humid area in Billwerder, where warmth and moisture come straight out of the ground when the sun shines. Although it is warmer and more humid directly on the ground in Hyytiälä Forest than above it, this has less direct effect on thermals. In some places, even a slight inversion forms in the shade of the trees. The thermals seem to come more from the warm and humid zone above the trees.

This zone enables a special type of detachment that only exists above forests and could explain the morning and evening thermals that are often observed there. It also happens when solar radiation is weak and the ground is not very warm: dry-cold downdrafts falling from above, acting as mass balancers for other updrafts, push under the air near the ground, lifting it and thus triggering the next updraft (arrows in Fig. 4).

Recognising downdrafts is relatively simple. It starts with the temperature curves "bulging down" because the air coming from above is colder than the over-adiabatic layer into which it falls. At the same time, the downdraft is usually quite dry and initially appears dark yellow in the diagrams. Its arrival at the ground is clearly visible. There, too, a yellow stripe pushes itself between the blue of the surrounding air. Immediately afterwards, the temperature and density lines of the air near the ground rise because the downdraft has pushed itself underneath them. Waves are literally triggered in the over-adiabatic layer. It's a bit like dropping a stone into a puddle.

The temperature of the ambient air drops over-adiabatically up to approx. 60 m altitude, i.e. by more than -1 °C/100 m. If the fall wind lifts the air above the forest, it is quickly warmer locally and, above all, more humid than the surroundings and can easily lift off. The differences between thermals due to ground heat and thermals due to lifted forest air are shown in **Fig. 5**. Of course, thermals due to warm forest soil also occur in Hyytiälä on other days and in other weather conditions.

The measurements also make it clear why we sometimes have the impression of finding a last updraft on approach, of all things, although we have searched in vain before. No sooner we radio our intention to land we are suddenly lifted upwards. This is because the over-adiabatic layer is constantly in motion, swinging up and down. Not every swing turns into an updraft, sometimes it only goes up a few dozen metres. Since the climb rates are similar to those of a thermal, one might mistakenly want to start a circle. But of course that would not be a good idea.

## Comparison with logger data from a glider

Finally, a special example from 29 May 2019. After an eight-hour cross-country flight, a two-seater from Boberg arrives back over its home airfield in the late afternoon (**Fig. 6**). Obviously the pilot team does not want to land immediately and circles a little near the airfield. But the thermals are already becoming unreliable. This also applies to the updraft directly by the weather mast, in which they want to gain altitude at 16:32 UTC. This thermal simply stops from one moment to the next, and the pilots have to try their luck elsewhere. But the following upwinds nearby are not very promising either. They find the last thermal at 16:57 UTC.

**Fig. 7** shows the associated measurement data [OLC1]. It is not quite as clear as in the previous pictures. Without going into too much detail, a few things spontaneously catch the eye. The thermals near the mast appear in the data. The updraft at 16:30 UTC is rising at about 2 m/s. This agrees with the logger readings from the OLC, taking into account the glider's own sinking. The next thermal does not occur until 17:04 UTC. There is actually "dead air" between the two thermals. The first thermal stops spontaneously, and it takes half an hour for another short two-minute thermal to form. The afternoon sun is no longer strong enough to warm the ground and evaporate the soil moisture. After the updraft has transported the last of the heat and moisture upwards at 16:30 UTC, the ground-level reservoir is exhausted for some time. The fact that there is still enough for a short rise at 17:04 UTC is probably more of a coincidence. The warm, humid air cushion that then takes off is no longer particularly productive. After that, thermals are finished and the pilots return to Boberg.

## Conclusion

The mechanisms leading to thermal detachment are different over grasslands than over forested areas. Over grasslands, warm, moist air bubbles typically form on the ground, are pushed by downdrafts falling from higher altitudes and subsequently rise due to their lower density compared to the surrounding air. Above forests, on the other hand, there is a relatively broad and homogeneous layer of warm, moist air above the trees. Downdrafts pass through this layer and cause it to rise locally. Since the local rise then has a lower density than the surrounding environment, a thermal bubble is released.

These effects become visible when data from boundary layer measurement masts are processed in such a way that height dependencies are eliminated. This can be done by transforming temperature and humidity measurement data to potential temperatures, mixing ratio and potential density.

**Figures**

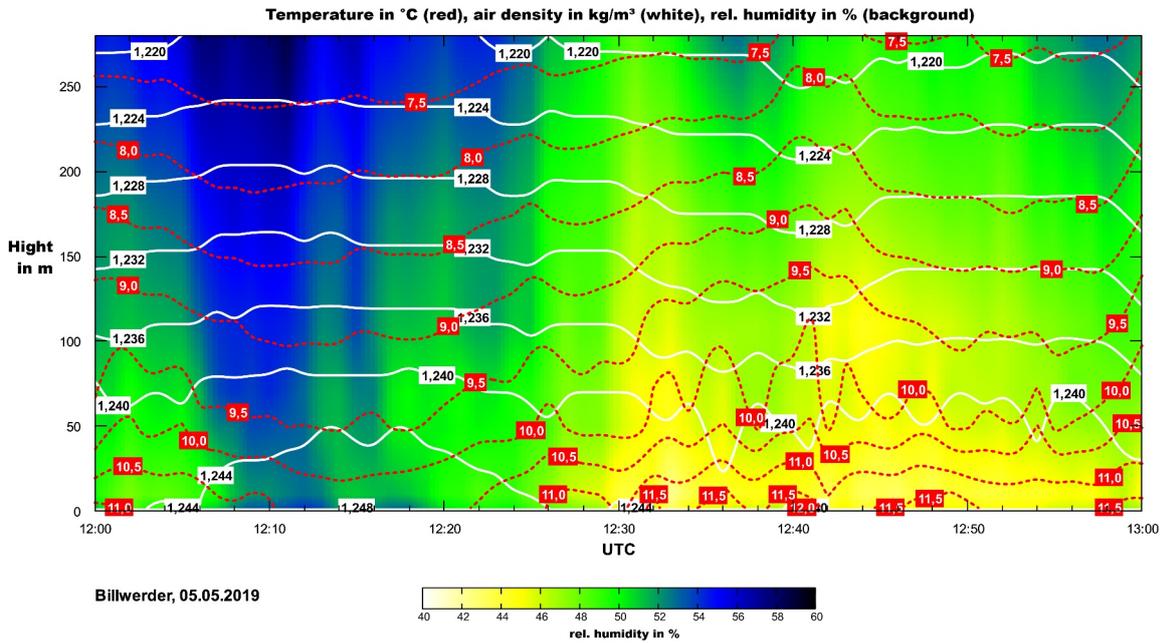

Fig. 1: Measured values of the boundary layer measuring mast in Hamburg-Billwerder.

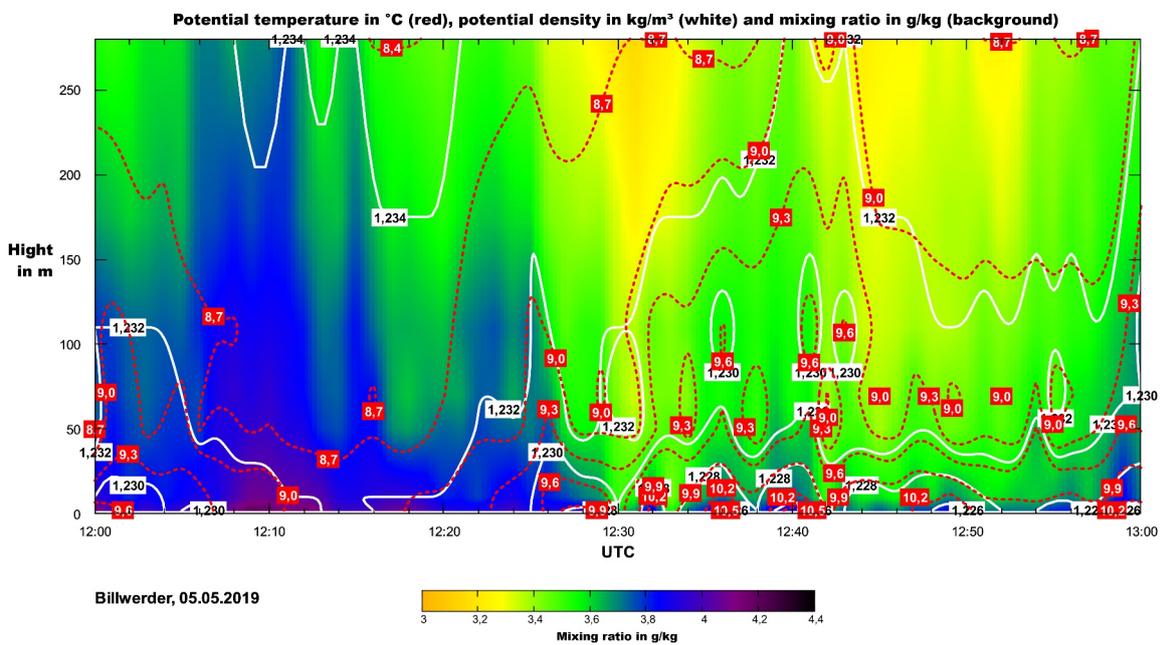

Fig. 2: Processed measurement data show updrafts and downdrafts.

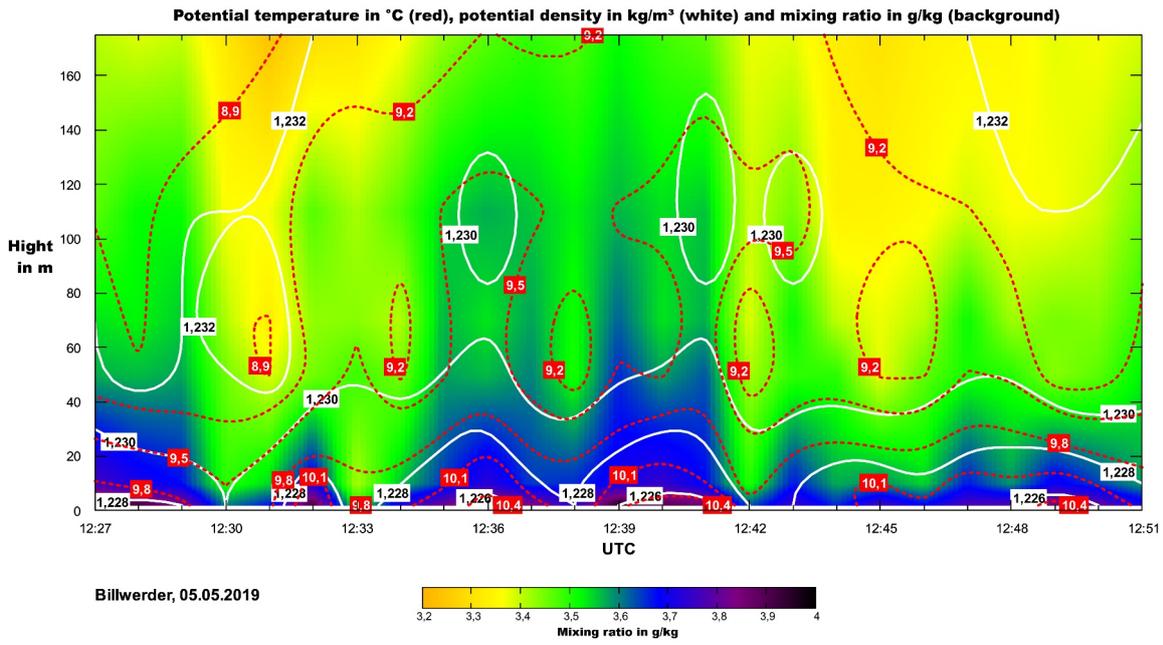

Fig. 3: The ground-level detachment in detail.

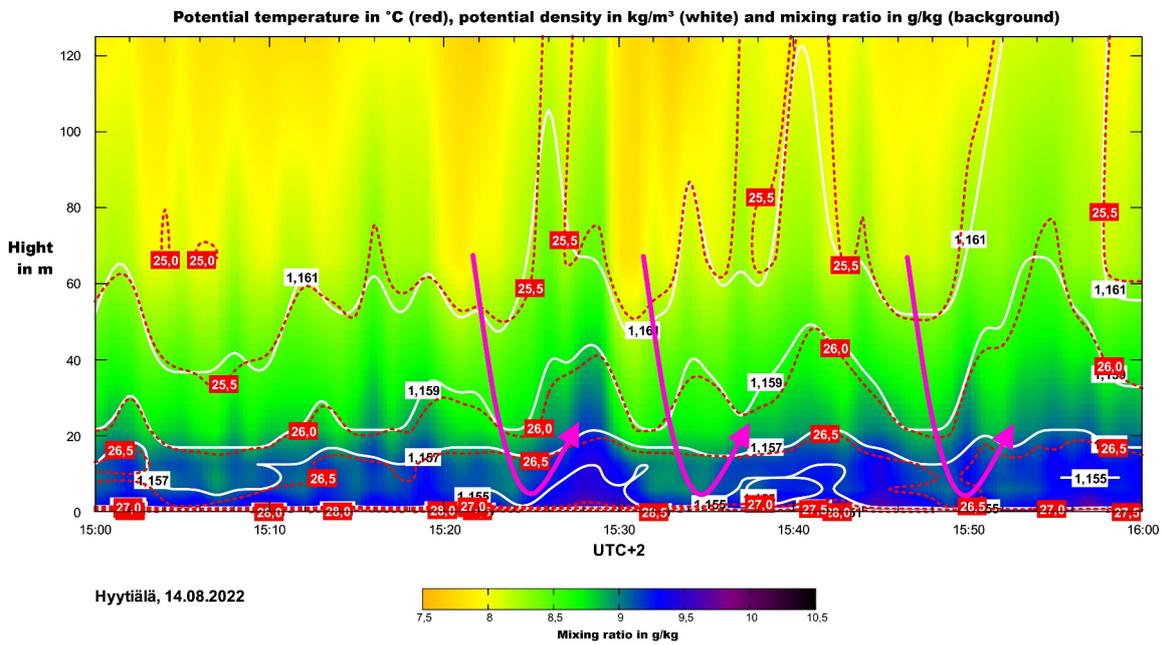

Fig. 4: Processed measurement data from the Hyytiälä pine forest in Finland.

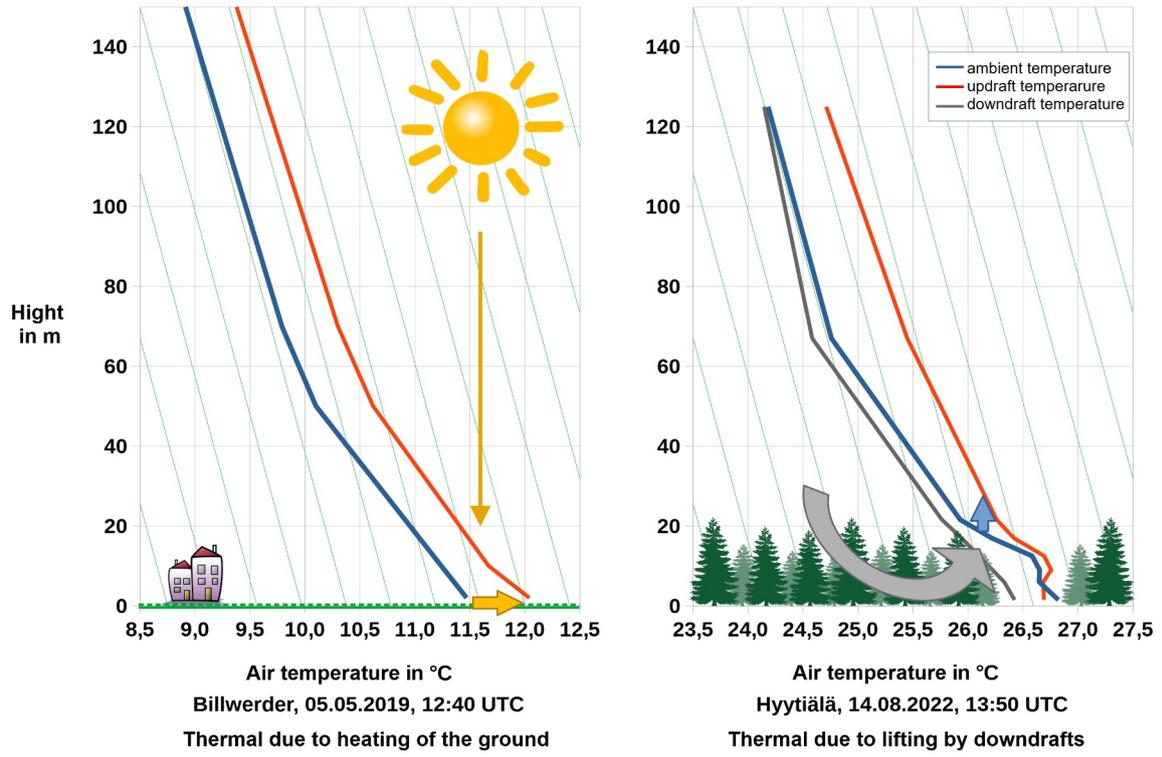

Fig. 5: Air temperatures in thermals due to ground heat (Billwerder) and thermals due to lifted forest air (Hyytiälä).

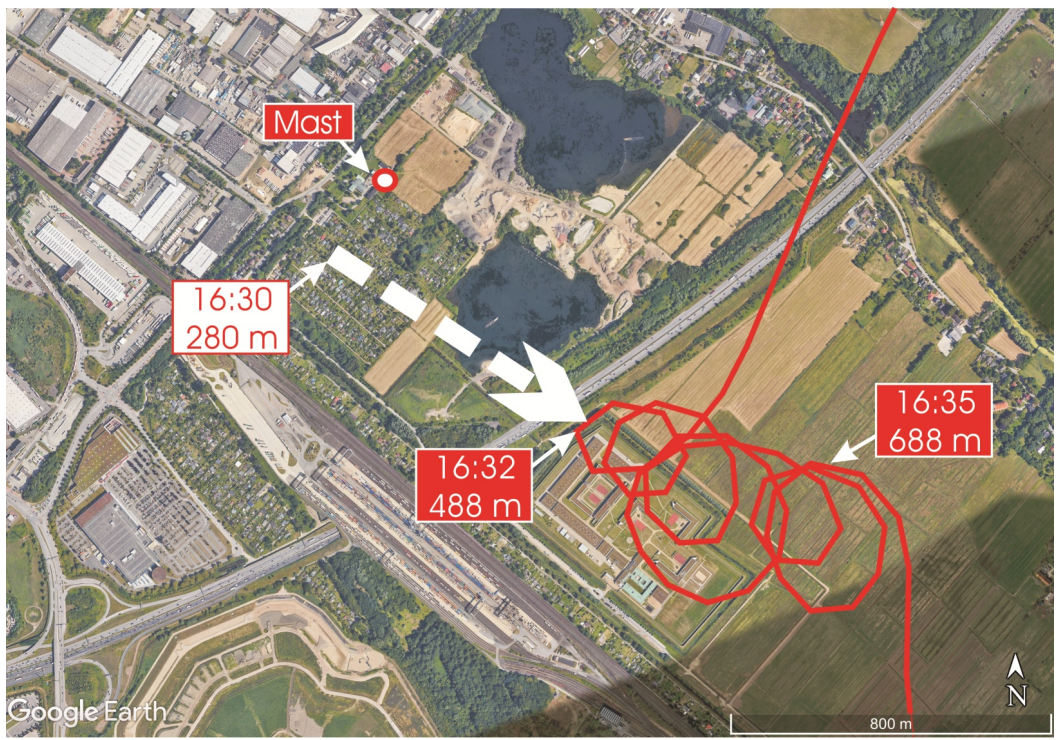

Fig. 6: Circling over Billwerder on May 29, 2019.

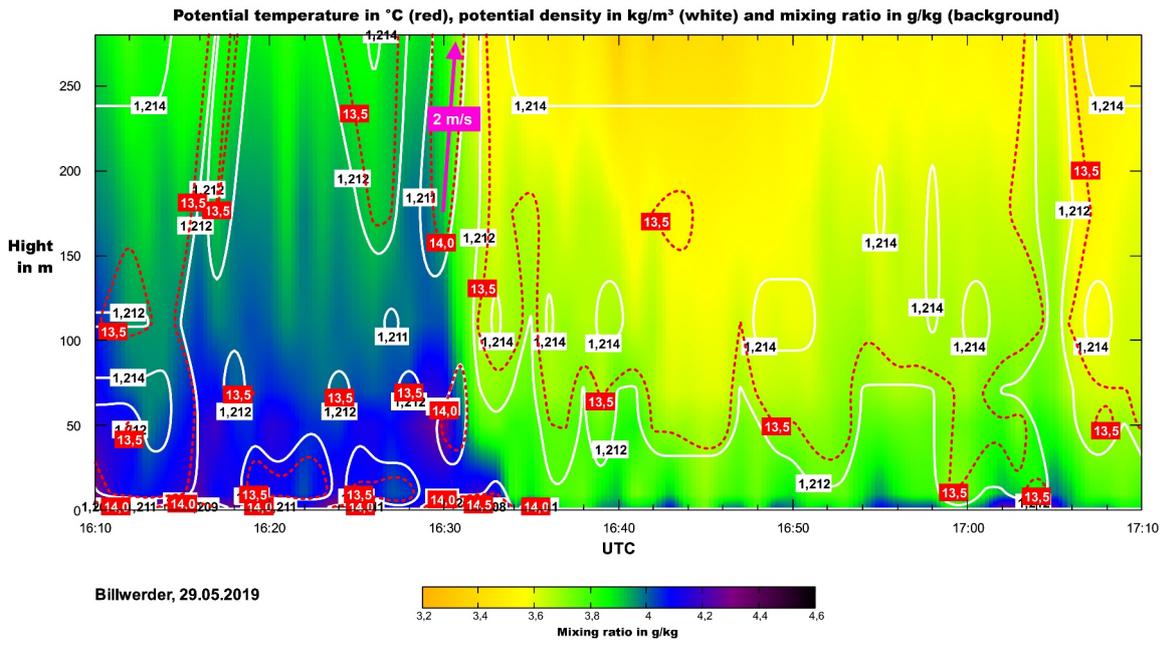

Fig. 7: Thermal end in Billwerder with climb rate of the last usable updraft.